\documentclass[final]{svjour3}
\usepackage{graphicx}
\usepackage{rotating}
\usepackage{amssymb}
\usepackage{mathptmx}
\usepackage[numbers]{natbib}

\usepackage{ascmac}
\usepackage{amsmath}
\usepackage{txfonts}

\makeatletter
\journalname{Journal of Low Temperature Physics}

\newcommand{\bs}{\boldsymbol}
\newcommand{\lef}{\left}
\newcommand{\rig}{\right}
\newcommand{\up}{\uparrow}
\newcommand{\down}{\downarrow}

\newcommand{\vep}{\varepsilon}

\def\lesssim{\ \raise.3ex\hbox{$<$}\kern-0.8em\lower.7ex\hbox{$\sim$}\ }
\def\gesim{\ \raise.3ex\hbox{$>$}\kern-0.8em\lower.7ex\hbox{$\sim$}\ }

\bibpunct{}{}{,}{s}{}{,}
\begin{document}
\newcommand{\hdblarrow}{H\makebox[0.9ex][l]{$\downdownarrows$}-}
\title{Strong Coupling Effects on the Specific Heat of an Ultracold Fermi Gas in the Unitarity Limit}
\author{P. van Wyk$^1$ \and H. Tajima$^1$ \and R. Hanai$^1$\and Y. Ohashi$^1$}
\institute{1:Department of Science and Technology, Keio University, Japan}
\date{23.07.2015}
\maketitle
\begin{abstract}
We investigate strong-coupling corrections to the specific heat $C_V$ in the normal state of an ultracold Fermi gas in the BCS-BEC crossover region. A recent experiment on a $^6$Li unitary Fermi gas [M. J. H. Ku, {\it et. al.}, Science {\bf 335}, 563 (2012)] shows that $C_V$ is remarkably amplified near the superfluid phase transition temperature $T_{\rm c}$, being similar to the well-known $\lambda$-structure observed in liquid $^4$He. Including pairing fluctuations within the framework of the strong-coupling theory developed by Nozi\`eres and Schmitt-Rink, we show that strong pairing fluctuations are sufficient to explain the anomalous behavior of $C_V$ observed in a $^6$Li unitary Fermi gas near $T_{\rm c}$. We also show that there is no contribution from {\it stable} preformed Cooper pairs to $C_V$ at the unitarity. This indicates that the origin of the observed anomaly is fundamentally different from the case of liquid $^{4}$He, where {\it stable} $^4$He Bose atoms induce the $\lambda$-structure in $C_V$ near the superfluid instability. Instead, the origin is the suppression of the entropy $S$, near $T_{\rm c}$, due to the increase of {\it metastable} preformed Cooper pairs. Our results indicate that the specific heat is a useful quantity to study the effects of pairing fluctuations on the thermodynamic properties of an ultracold Fermi gas in the BCS-BEC crossover region.

\end{abstract}
\par
\section{Introduction}
\par
Ultracold Fermi gases have recently gathered much theoretical and experimental attention as highly controllable many-body systems\cite{ref_int_1,ref_int_2,ref_int_3,ref_int_4,ref_int_5,ref_int_6,ref_int_7,ref_int_8}. A tunable pairing interaction associated with a Feshbach resonance\cite{ref_int_6} has realized the so-called BCS (Bardeen-Cooper-Schrieffer)-BEC (Bose Einstein condensation) crossover phenomenon, where we can study a Fermi superfluid and a Bose superfluid in a unified manner. In the BCS-BEC crossover region, the system is dominated by strong pairing fluctuations, so that the formation of preformed Cooper pairs, as well as their effects on physical properties, have extensively been discussed\cite{ref_int_9,ref_int_10,ref_int_11}. Recently various thermodynamic quantities have become accessible in the field of cold Fermi gas physics\cite{ref_int_12,ref_int_13,ref_int_14,ref_int_15}, allowing us to directly compare theoretical studies of the strong-coupling properties of an ultracold Fermi gas\cite{ref_int_10,ref_int_16,ref_int_17,ref_int_18} with experimental results. 
\par
Since the formation of preformed Cooper pairs due to a strong pairing interaction is expected to affect the entropy $S$ of the system, the specific heat (which is directly related to the change in entropy) is a useful thermodynamic quantity to study this key many-body phenomenon in the BCS-BEC crossover region. Indeed, a recent experiment on a $^6$Li unitary Fermi gas\cite{ref_int_18} observed that the specific heat at constant volume $C_V$ deviates from the linear temperature dependence known in a normal Fermi liquid\cite{ref_int_19}. In the normal state near $T_{\rm c}$, the observed $C_V$ exhibits an anomalous enhancement with decreasing temperature, which resembles the so-called $\lambda$-structure observed in liquid $^4$He\cite{ref_int_20}. Although this resemblance infers that some form of Bose excitations are responsible for the anomalous behavior of $C_V$ in a $^6$Li unitary Fermi gas near $T_{\rm c}$, it is unclear to what extent a unitary Fermi gas with strong pairing fluctuations can be regarded as an interacting Bose system, like liquid $^4$He. 
\par
In this paper, we investigate the specific heat at constant volume $C_V$, and effects of pairing fluctuations in the BCS-BEC crossover regime of an ultracold Fermi gas, above the superfluid phase transition temperature $T_{\rm c}$. Including pairing fluctuations within the framework developed by Nozi\`eres and Schmitt-Rink (NSR)\cite{ref_int_21}, we show that strong-coupling corrections to $C_V$, near $T_{\rm c}$, in the unitarity limit are sufficient to explain the anomalous enhancement of $C_V$ that has recently been observed in a $^6$Li Fermi gas. We also clarify that this phenomenon originates from pairing fluctuations leading to the formation of {\it metastable} preformed pairs, which is quite different from the case of liquid $^4$He, where {\it stable} $^4$He Bose atoms induce the $\lambda$-structure in the temperature dependence of $C_V$ near the superfluid instability\cite{ref_int_20}. Throughout this paper we take $\hbar=k_{\rm B}=1$, and the system volume $V$ is taken to be unity, for simplicity.     
\par
\par
\section{Formulation}
\par
We consider a two-component Fermi gas, described by the BCS Hamiltonian,
\begin{equation}
H= \sum_{\bs{p},\sigma}\xi_{\bs p}c^{\dagger}_{\bs{p},\sigma}c_{\bs{p},\sigma}
-U\sum_{\bs{p},\bs{p}',\bs{q}}
c^{\dagger}_{\bs{p}+\bs{q}/2,\up}
c^{\dagger}_{-\bs{p}+\bs{q}/2,\down}
c_{-\bs{p}'+\bs{q}/2,\down}
c_{\bs{p}'+\bs{q}/2,\up},
\label{eq.1} 
\end{equation}    
where $c_{\bs{p},\sigma}$ is the annihilation operator of a Fermi atom with pseudospin $\sigma=\uparrow,\downarrow$, describing two atomic hyperfine states. $\xi_{\bs p}=\varepsilon_{\bs p}-\mu=p^2/(2m)-\mu$ is the kinetic energy, measured from the Fermi chemical potential $\mu$, where $m$ is an atomic mass. The pairing interaction $-U$ ($<0)$ is assumed to be tunable by adjusting the threshold energy of a Feshbach resonance\cite{ref_int_6}. As usual, we measure the interaction strength in terms of the inverse $s$-wave scattering length $a_s^{-1}$, which is related to $-U$ through
\begin{equation}
\frac{4\pi a_{s}}{m}=\frac{-U}{1-U\sum_{\bs{p}}\frac{1}{2\vep_{\bs{p}}}}.
\label{eq.2}
\end{equation}
\par
\begin{figure}[t]
\begin{center}
\includegraphics[width=0.6\textwidth]{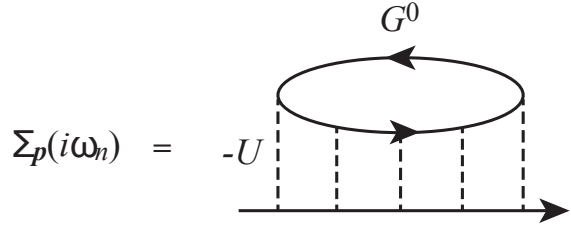}
\caption{Self-energy $\Sigma_{\bs q}(i\omega_n)$ in the NSR Green's function in Eq. (\ref{eq.3}). The solid line and dashed line are the non-interacting Green's function $G^0$, and the pairing interaction $-U$, respectively.}
\label{fig1}       
\end{center}
\end{figure}
\par
We include fluctuations in the Cooper channel within the framework of the strong-coupling theory developed by Nozi\`eres and Schmitt-Rink (NSR)\cite{ref_int_21}. In the Green's function formalism, the NSR theory is described by the single-particle thermal Green's function,
\begin{equation}
G_{\bs p}(i\omega_n)=G^{0}_{\bs p}(i\omega_n)
+
G^0_{\bs{p}}(i\omega_n)
\Sigma_{\bs p}(i\omega_n)
G^0_{\bs{p}}(i\omega_n),
\label{eq.3}
\end{equation}
where $\omega_{n}$ is the fermion Matsubara frequency, and $G^{0}_{\bs{p}}(i\omega_n)^{-1}=i\omega_n-\xi_{\bs p}$ is the Green's function for a free Fermi particle. The NSR self-energy $\Sigma_{\bs p}(i\omega_n)$, which describes strong-coupling corrections to single-particle excitations, is diagrammatically drawn as Fig. \ref{fig1}, which gives,
\begin{equation}
\Sigma_{\bs p}(i\omega_{n})=T\sum_{\bs{q},i\nu_{n}}
\Gamma_{\bs q}(i\nu_{n})G^{0}_{\bs{p}-\bs{q}}(i\omega_{n}-i\nu_{n}).
\label{eq.4} 
\end{equation}
Here, $\nu_n$ is the boson Matsubara frequency. The NSR particle-particle scattering matrix,
\begin{equation}
\Gamma_{\bs q}(i\nu_{n}) = 
{1
\over
{m \over 4\pi a_s}+
\left[
\Pi_{\bs q}(i\nu_n)-\sum_{\bs p}{1 \over 2\varepsilon_{\bs p}}
\right]
},			
\label{eq.5}
\end{equation}
describes fluctuations in the Cooper channel, where
\begin{equation}
\Pi_{\bs q}(i\nu_{n}) = -\sum_{\bs{p}}\frac{1-f(\xi_{\bs{p}+\bs{q}/2})-f(\xi_{\bs{p}-\bs{q}/2})}{i\nu_{n}-\xi_{\bs{p}+\bs{q}/2}-\xi_{\bs{p}-\bs{q}/2}}
\label{eq.6}
\end{equation}   
is the lowest order pair-correlation function (where $f(x)$ is the Fermi distribution function). We briefly note that, although the pair correlation function $\Pi_{\bs{q}}(i\nu_{n})$ has an ultraviolet divergence, this singularity is renormalized into the $s$-wave scattering length in Eq. (\ref{eq.2}). 
\par
The specific heat at constant volume $C_V$ is calculated from the formula, 
\begin{equation}
C_{V}=\lef({\partial E \over  \partial T}\rig)_{V,N}.
  \label{eq.7} 
\end{equation}
Here, the NSR internal energy $E$ is given by
\begin{eqnarray}
E &=& 2T\sum_{\bs{p},i\omega_n}
\left[
\varepsilon_{\bs{p}}+\frac{1}{2}\Sigma_{\bs{p}}(i\omega_{n})
\right]
G_{\bs p}(i\omega_n)
\nonumber\\
&\simeq&
-T\sum_{\bs{q},i\nu_{n}}\Gamma_{\bs q}(i\nu_n)
\left[
T\frac{\partial}{\partial{T}}\Pi_{\bs q}(i\nu_n)
+\mu\frac{\partial}{\partial{\mu}}\Pi_{\bs q}(i\nu_n)
\right],
\label{eq.8} 
\end{eqnarray}
where we have retained only terms to $O(\Sigma_{\bs p}(i\omega_n))$, so as to be consistent with the NSR Green's function in Eq. (\ref{eq.3}). In evaluating (\ref{eq.7}) and (\ref{eq.8}), we need the value of the chemical potential $\mu(T)$, which is obtained from the equation for the number $N$ of Fermi atoms,
\begin{align}
N= 2T\sum_{\bs{p},i\omega_{n}}G_{\bs{p}}(i\omega_n). 
\label{eq.9} 
\end{align}
Equation (\ref{eq.9}) is also used in self-consistently determining the superfluid phase transition temperature $T_{\rm c}$ from the $T_{\rm c}$-equation\cite{ref_int_21},
\begin{equation}
{m \over 4\pi a_s}=-
\sum_{\bs p}
\left[
{1 \over 2\xi_{\bs p}}\tanh{\xi_{\bs p} \over 2T_{\rm c}}
-{1 \over 2\varepsilon_{\bs p}}
\right],
\label{eq.10}
\end{equation}
which is obtained from the Thouless criterion\cite{ref_int_22}, which states that the superfluid instability occurs when the particle-particle scattering matrix $\Gamma_{\bs q}(i\nu_n)$ has a pole at ${\bs q}={\bs 0}$ and $\nu_n=0$. 
\par
In our numerical calculations, we tune the interaction strength by adjusting the value of the dimensionless parameter $\lef(k_{\rm F} a_s\rig)^{-1}$. We briefly note that, in the unitarity limit ($\lef(k_{\rm F} a_s\rig)^{-1}=0$), the interaction effect formally disappears in Eqs. (\ref{eq.5}) and (\ref{eq.10}), in the sense that the factor $m/(4\pi a_s)$ in these equations vanishes.

\begin{figure}[t]
\begin{center}
\includegraphics[width=1.0\linewidth,keepaspectratio]{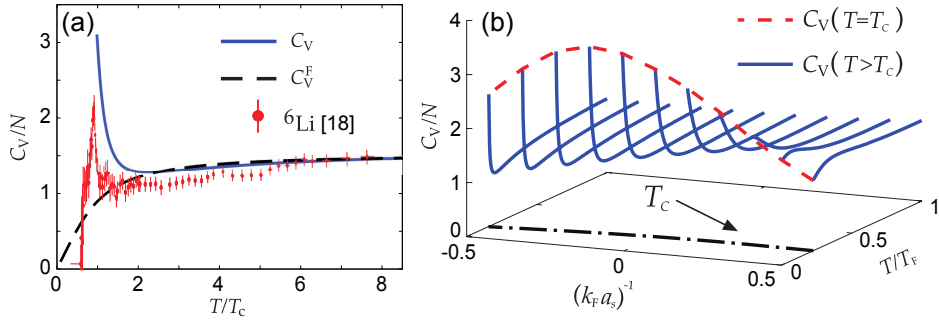}
\end{center}
\caption{(Color online) (a) Comparison of our result with the recent experiment on a $^{6}$Li unitary Fermi gas\cite{ref_int_18}. $C_V^{\rm F}$ shows the specific in a free Fermi gas. (b) Calculated specific heat $C_V$ in the BCS-BEC crossover regime of an ultracold Fermi gas above $T_{c}$\cite{note_2}.}
\label{fig.2}
\end{figure}
\par
\section{Specific heat in the BCS-BEC crossover region above $T_{\rm c}$}
\par
Figure \ref{fig.2}(a) shows the specific heat $C_V$ in a unitary Fermi gas above $T_{\rm c}$. Starting from the high temperature region, $C_V$ decreases from the classical result $C_V^{\rm cl}=(3/2)N$, with decreasing temperature. Since this high temperature behavior is well described by the case of a free Fermi gas ($C_V^{\rm F}$), this temperature dependence may be simply attributed to the Fermi statistical effect. While the free Fermi gas result $C_V^{\rm F}$ monotonically decreases with decreasing temperature, the specific heat $C_V$ at the unitarity is enhanced in the low temperature region, $T_{\rm c}\le T\lesssim 2T_{\rm c}$. As seen in Fig. \ref{fig.2}(a), this enhancement semi-quantitatively agrees with the recent experiment on a $^6$Li Fermi gas\cite{ref_int_16,note_1}. Since the physical properties of a unitary Fermi gas are known to be dominated by strong pairing fluctuations near $T_{\rm c}$\cite{ref_int_8,ref_int_9,ref_int_10,ref_int_16,ref_int_17,ref_int_21,ref_int_23}, this anomaly in $C_V$ is also considered as a strong-coupling phenomenon associated with strong fluctuations in the Cooper channel. Indeed, as shown in Fig. \ref{fig.2}(b), this enhancement is remarkable in the unitary regime.
\par
At a glance, the anomalous enhancement of the specific heat $C_V$ seen in Fig. \ref{fig.2}(a) resembles the case of liquid $^4$He\cite{ref_int_20}, where stable atoms constituting this Bose liquid are responsible for the famous $\lambda$-structure in the temperature dependence of the specific heat near the superfluid instability. Since pairing fluctuations in a unitary Fermi gas are accompanied by preformed Cooper pairs, it is an interesting problem whether or not the observed enhancement of $C_V$ can be understood by regarding a unitary Fermi gas as a ``Bose gas" of stable preformed Cooper pairs.
\par
To examine this, it is convenient to divide the number equation (\ref{eq.9}) into the sum of the free fermion part, 
\begin{equation}
N_{\rm F}= 2T\sum_{\bs{p},i\omega_{n}}G^0_{\bs{p}}(i\omega_n)=2\sum_{\bs{p}}f(\xi_{\bs{p}}),
\label{eq.10b}
\end{equation}
and the fluctuation contribution coming from the second term in Eq. (\ref{eq.3}), 
\begin{eqnarray}
N_{\rm FL}
=
- T\sum_{{\bs q},i\nu_n}
\Gamma_{\bs q}(i\nu_n)
{\partial \over \partial\mu}\Pi_{\bs q}(i\nu_n)
=
\int_{-\infty}^{\infty}{\rm d}\omega n_{\rm B}(\omega)\rho_{\rm B}(\omega).
\label{eq.11}
\end{eqnarray}
Here, $n_{\rm B}(\omega)$ is the Bose distribution function, and 
\begin{equation}
\rho_{\rm B}(\omega)=\sum_{\bs{q}}\frac{-1}{\pi}\text{Im}
\left[
\Gamma_{\bs q}(i\nu_n\to\omega+i\delta)
{\partial \over \partial\mu}\Pi_{\bs q}(i\nu_n\to\omega+i\delta)
\right]
\label{eq.12}
\end{equation}
can be interpreted as the ``molecular density of states" (where $\delta$ is an infinitesimally small positive number). Equation (\ref{eq.11}) implies that $N_{\rm FL}$ is related to the bosonic character of the system. Evaluating the contribution from {\it real} poles ($\Omega_{\bs q}$'s) of the analytic-continued particle-particle scattering matrix $\Gamma_{\bs q}(i\nu_n\to\omega+i\delta)$ to Eq. (\ref{eq.11}) ($\equiv N_{\rm ST}$), one has\cite{ref_int_21,ref_int_22}
\begin{equation}
N_{\text{ST}}  
= \sum_{\Omega_{\bs q}}
\left[
\frac{\frac{\partial}{\partial{\mu}}\Pi_{\bs q}(\Omega_{\bs{q}})}
{\frac{\partial}{\partial{\Omega_{\bs{q}}}}\Pi_{\bs q}(\Omega_{\bs{q}})}
\right]
n_{\rm B}(\Omega_{\bs q}),
\label{eq.13} 
\end{equation}
where the summation is taken over real poles ($\Omega_{\bs q}$'s). Apart from the prefactor of the Bose distribution function in Eq. (\ref{eq.13}), $N_{\rm ST}$ is found to be directly related to the number of stable molecular bosons having the dispersion $\Omega_{\bs q}$. In the extreme BEC limit (where the system is well described by a gas of tightly bound molecules), the prefactor is reduced to two, and the molecular dispersion becomes $\Omega_{\bs q}=q^2/2M$ (where $M=2m$), as expected. To conclude, the fluctuation contribution $N_{\rm FL}=N_{\rm ST}+N_{\rm SC}$ can be decomposed into twice the number of stable preformed Cooper pair molecules $N_{\rm ST}$ and the so-called scattering part $N_{\rm SC}$\cite{ref_int_21,ref_int_22}, the latter of which physically describes the number of {\it metastable} preformed pairs with a finite lifetime.
\par
\begin{figure}[t]
\begin{center}
\includegraphics[width=0.6\linewidth,keepaspectratio]{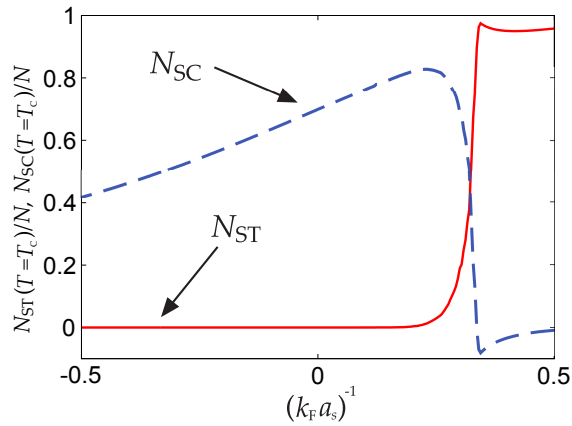}
\end{center}
\caption{(Color online) Calculated number of stable preformed pairs $N_{\rm ST}(T=T_{\rm c})$, as well as and contribution of scattering states $N_{\rm SC}(T=T_{\rm c})$. The fluctuation contribution $N_{\rm FL}$ is given by the sum of the two as $N_{\rm FL}=N_{\rm SC}+N_{\rm ST}$ \cite{note_2}.}
\label{fig.3}
\end{figure}
\par
As shown in Fig. \ref{fig.3}, while the fluctuation contribution $N_{\rm FL}$ at $T_{\rm c}$ is dominated by stable molecules ($N_{\rm ST}$) in the strong-coupling BEC regime, they are absent in the unitarity limit\cite{note_2}. This indicates that, although the enhanced specific heat $C_V(T\sim T_{\rm c})$ shown in Fig. \ref{fig.2}(a) is similar to the $\lambda$-structure in the case of liquid $^4$He, their origins are different from each other. While {\it stable} $^4$He atoms are responsible for this anomaly in a liquid $^4$He, the suppression of the entropy due to the formation of {\it metastable} preformed pairs, with decreasing temperature, leads to the amplification of the specific heat near $T_{\rm c}$ in a unitary Fermi gas, through the thermodynamic relation,
\par
\begin{equation}
C_{V}=T\lef(\frac{\partial S}{\partial T}\rig)_{V,N}.
\end{equation}       
\par
We see in Fig. \ref{fig.3} that the number $N_{\rm SC}$ of metastable Cooper pairs at $T_{\rm c}$ first increases, as one approaches the unitarity regime from the weak-coupling BCS side. Since the increase of this fluctuation contribution suppresses the entropy $S(T\sim T_{\rm c})$, the specific heat $C_V$ is enhanced near $T_{\rm c}$, as shown in Fig. \ref{fig.2}(b). Such a fluctuation effect on the entropy $S$ is, however, suppressed in the strong-coupling BEC regime, because metastable preformed Cooper pairs are replaced by stable tightly bound molecules with a finite binding energy\cite{ref_int_21} $E_{\rm bind}=1/{ma_{s}^2}$. (See Fig. \ref{fig.3}.) This is the reason for the non-monotonic behaviour of $C_V(T=T_{\rm c})$ with respect to the interaction strength seen in Fig. \ref{fig.2}(b). 
\par
In the strong-coupling BEC regime, the system is well described by a gas of $N_{\rm B}=N/2$ tightly bound molecules with the molecular mass $M=2m$. As a result, $C_V$ increases with decreasing the temperature near $T_{\rm c}$ when $(k_{\rm F}a_s)^{-1}\gesim 0.8$, as in the case of an ideal Bose gas (although we do not explicitly show this in this paper). In the BEC limit, $C_V(T=T_{\rm c})$ just equals the specific heat $C^{\rm BEC}_V\simeq 1.93N_{\rm B}$ of an ideal molecular Bose gas at the BEC phase transition temperature\cite{Fetter}.
\par   
\par
\section{Summary}
\par
To summarise, motivated by the prospective usefulness of the specific heat $C_V$ as a probe to study effects of strong pairing fluctuations in an ultracold Fermi gas, we have discussed strong coupling corrections to this thermodynamic quantity in the BCS-BEC crossover region. Within the framework of the strong-coupling theory developed by Nozi\`eres and Schmitt-Rink, we showed that $C_V$ is enhanced in the unitary regime near $T_{\rm c}$. We also showed that this anomalous enhancement is sufficient to explain the recent experiment on a $^6$Li unitary Fermi gas, where the observed specific heat is amplified with decreasing the temperature near $T_{\rm c}$.
\par
Although this behaviour looks similar to the $\lambda$-structure of the temperature dependence of the specific heat in a liquid $^4$He near the superfluid phase transition temperature, we pointed out that the origin of the former is fundamentally different from the latter. Indeed, we found that there are no stable preformed Cooper pairs at the unitarity, which is quite different from the latter case, where $^4$He atoms are always stable. Instead, the bosonic character of a unitary Fermi gas near $T_{\rm c}$ is dominated by the increase of metastable preformed pairs as the temperature is lowered. This naturally suppresses the entropy near $T_{\rm c}$, leading to the amplification of $C_{V}$. Our result indicates that the anomalous amplification of the specific heat observed in a $^6$Li unitary Fermi gas can be fully explained by fluctuating Fermi pairs, without the inclusion of stable bosons.
\par
In this paper, we have only treated the normal state. Since the specific heat has also been observed in the superfluid phase, it is important to extend the present theory to the region below $T_{\rm c}$. In addition, since a real Fermi gas is trapped in a harmonic potential, effects of spatial inhomogeneity on the specific heat is also a crucial future problem. Since the specific heat is a fundamental thermodynamic quantity, our results would contribute to the further understanding of thermodynamic properties of an ultracold Fermi gas in the BCS-BEC crossover region.
\par
\par
\begin{acknowledgements}
We thank D. Inotani and M. Matsumoto for discussions. This work was supported by the KiPAS project in Keio university. Y.O was supported by Grant-in-Aid for Scientific Research from MEXT and JSPS in Japan (No.25400418, No.15H00840).
\end{acknowledgements}
\par


\begin{thebibliography}{99}
\bibitem{note_1} We note that the difference between the calculated $C_{V}$ and experimental result seen in Fig. \ref{fig.2}(a) could be due to an inherent uncertainty in the temperature resolution of the experiment, because the temperature cannot be directly measured\cite{ref_int_18}. The inclusion of such effects to the calculation of $C_{V}$ is left as our future work.
\bibitem{note_2} We briefly note that $N_{\rm SC}$ in Fig. \ref{fig.3} is slightly negative in the strong-coupling regime when $(k_{\rm F}a_s)^{-1}\gesim 0.3$. This is because, strictly speaking, the molecular picture is not exact, when preformed Cooper pairs have a finite lifetime. Thus, $N_{\rm SC}$ involves fluctuation effects that cannot be fully understood within the simple particle picture\cite{ref_int_23}.
\end{thebibliography}
\end{document}